\documentclass[prd,
twocolumn,superscriptaddress,preprintnumbers,nofootinbib]{revtex4}
\usepackage{graphicx}
\usepackage{epsfig}
\usepackage{bm}
\usepackage{latexsym,amssymb,amsmath,amssymb,wasysym,float}

\usepackage{color}

\usepackage{scrextend}
\usepackage{mathrsfs}
\usepackage{enumitem}

\newcommand{\postscript}[2]{\setlength{\epsfxsize}{#2\hsize}
   \centerline{\epsfbox{#1}}}

\usepackage[usenames,dvipsnames]{xcolor}
\definecolor{orange}{cmyk}{0,0.5,1,0}
\definecolor{rossoCP3}{cmyk}{0,.88,.77,.40}
\definecolor{graa}{rgb}{0.8,0.8,0.8}
\definecolor{blaa}{rgb}{0.2,0.2,0.6}

\begin{document}

\preprint{MPP-2025-45}
\preprint{LMU-ASC 06/25}

\title{\color{rossoCP3} $\bm{S}$-dual Quintessence, the Swampland, and
  the DESI DR2 Results}
\author{\bf Luis A. Anchordoqui}

\affiliation{Department of Physics and Astronomy,  Lehman College, City University of
  New York, NY 10468, USA
}

\affiliation{Department of Physics,
 Graduate Center,  City University of
  New York,  NY 10016, USA
}

\affiliation{Department of Astrophysics,
 American Museum of Natural History, NY
 10024, USA
}

\author{\bf Ignatios Antoniadis}

\affiliation{High Energy Physics Research Unit, Faculty of Science, Chulalongkorn University, Bangkok 1030, Thailand}

\affiliation{Laboratoire de Physique Th\'eorique et Hautes \'Energies
  - LPTHE \\
Sorbonne Universit\'e, CNRS, 4 Place Jussieu, 75005 Paris, France
}

\author{\bf Dieter\nolinebreak~L\"ust}

\affiliation{Max--Planck--Institut f\"ur Physik,  
 Werner--Heisenberg--Institut,
80805 M\"unchen, Germany
}

\affiliation{Arnold Sommerfeld Center for Theoretical Physics, 
Ludwig-Maximilians-Universit\"at M\"unchen,
80333 M\"unchen, Germany
}

\begin{abstract}
  \noindent We propose a dark energy model in which a quintessence
  field 
  $\phi$ rolls near the vicinity of a local maximum of its potential
  characterized by the simplest $S$ self-dual form $V(\phi) = \Lambda \ {\rm sech}
  (\sqrt{2} \, \phi/M_p)$, where $M_p$ is the reduced Planck mass and $\Lambda \sim 10^{-120} M_p^4$ is the cosmological
  constant. We confront the model with Swampland ideas and show that the $S$-dual potential is consistent
  with the distance conjecture, the de Sitter conjecture, and the trans-Planckian censorship conjecture. We also examine the compatibility of this phenomenological
    model with the intriguing DESI DR2 results and show that the shape of the
 $S$-dual potential is almost indistinguishable from the axion-like
 potential,  $V (\phi) = m_a^2 \ f_a^2 \ [ 1 + \cos(\phi/f_a)]$, with $m_a$ and $f_a$ parameters
 fitted by the DESI Collaboration to accommodate the DR2 data. The
 self-dual potential has the advantage that one starts at
 the self-dual point and this is a theoretical motivation, because as
the universe cools off the $\mathbb{Z}_2$ symmetry gets broken leading to a natural rolling away from the symmetric point.
\end{abstract}

\date{March 2025}
\maketitle

Dualities within gauge theories are
out of the ordinary because they connect
a strongly coupled field theory to a weakly coupled one. Accordingly, they are practical for evaluating a theory at strong coupling, where
perturbation theory breaks down, by translating it into its dual
description with a weak coupling constant. As a result, dualities point
to a single quantum system which has two classical limits. The $U(1)$
gauge theory on $\mathbb R^4$ is
a textbook
example, holding an electric-magnetic
duality symmetry that inverts the coupling constant and extends to an
action of $SL(2,\mathbb Z)$~\cite{Montonen:1977sn}. Several examples of $S$-duality in String
Theory have been investigated; see
e.g.~\cite{Font:1990gx,Sen:1994fa,Alvarez-Gaume:1996ohl,Gopakumar:2000na,
  Nekrasov:2004js,Argyres:2007cn,Gaiotto:2008ak,Dimofte:2011jd}. In
this Letter we first examine the consistency of potentials that are invariant under the $S$-duality constraint with ideas of the
Swampland
program~\cite{Vafa:2005ui}. Armed with our findings, we evaluate the
compatibility of $S$-dual quintessence models and the intriguing
results recently reported by the DESI
Collaboration~\cite{DESI:2025zpo,DESI:2025zgx}. Before
proceeding, we pause to note that we do not propose a direct
connection to a particular string
vacuum,
but simply think of the self-dual constraint as a relic of string
physics in the late time acceleration of our Universe.

The Swampland program seeks to set apart the space of effective field
theories (EFTs) coupled to gravity with a consistent UV completion~\cite{Vafa:2005ui}. Such a set of EFTs are planted in the landscape (i.e. the area
compatible with quantum gravity) while other EFTs are usually
relegated to an area called the swampland, which is incompatible with
quantum gravity. It is self-evident that the swampland is wider than
the landscape, and actually it surrounds the landscape. There are
several conjectures for fencing off the
swampland~\cite{Palti:2019pca,vanBeest:2021lhn,Agmon:2022thq}. The conjectures relevant to our investigation are
those related to effective scalar field theories canonically coupled
to gravity and endowed with a canonical kinetic term, which dominates the
energy density in the local universe; for an overview of the Swampland perspective and dark energy,
see~\cite{Vafa:2025nst}. Indeed, the following  conditions are
conjectured to hold for an EFT not to be downgraded to the swampland:
\begin{itemize}[noitemsep,topsep=0pt]
\item Distance conjecture: If a scalar field $\phi$ 
transverses a trans-Planckian range in the moduli space, a tower of string states becomes
light exponentially with increasing
distance~\cite{Ooguri:2006in}. As a consequence, the range traversed by scalar fields in field space is bounded by
\begin{equation}
  \Delta \phi < c_1 \, ,
\end{equation}
where $c_1 \sim {\cal O}(1)$ in reduced Planck units.
\item dS conjecture: The gradient of the potential $V$
  must satisfy either the lower bound,
\begin{equation}
|\nabla V| \geq c_2  \, V \,,
\label{c2}
\end{equation}
 or else (in the refined version of the dS conjecture) must satisfy
\begin{equation}
  {\rm min} (\nabla_i \nabla_jV ) \leq - c_3 \, V \,,
  \label{RdSC}
\end{equation}
where $c_2$ and
$c_3$ are positive order-one numbers in reduced Planck units and the left-hand side of (\ref{RdSC}) is the minimum eigenvalue of the
Hessian $\nabla_i \nabla_j V$  in an orthonormal frame~\cite{Obied:2018sgi,Garg:2018reu,Ooguri:2018wrx,Dvali:2018jhn,Dvali:2018fqu,Dvali:2017eba,Dvali:2014gua}.
\item TransPlanckian censorship conjecture (TCC): In any consistent theory of quantum gravity
any sub-Planckian fluctuation should remain quantum during any
cosmological expansion~\cite{Bedroya:2019snp}. This implies that for $d$-dimensional spacetimes in the asymptotic of
the field space, cosmologies driven by scalar fields, satisfy
\begin{equation}
  \left. \frac{|\nabla V|}{V}\right|_{\infty} \geq c_{\rm asymptotic} \, ,\end{equation}
where $c_{\rm asymptotic} = 2/
\sqrt{d - 2}$ in reduced Planck units~\cite{Bedroya:2019snp,Rudelius:2023mjy,vandeHeisteeg:2023uxj}.
\end{itemize}
These three conjectures set constraints on $S$-dual potentials. It is this
that we now turn to study.

For a real scalar field $\phi$, the $S$-duality symmetry
takes the form $\phi \to
-\phi$  (or analogously \mbox{$g\rightarrow 1/g$}, with $g\sim
e^{\phi/M_p}$, and where $M_p$ is the reduced Planck mass). For a complex scalar, the
$S$-duality group is extended to the modular group $SL(2, \mathbb
Z)$. The $S$-duality constraint forces a
particular functional form on the potential:
$f[\cosh(\phi/M_p)]$~\cite{Anchordoqui:2014uua,Anchordoqui:2021eox}. Herein, we
examine the simplest $S$ self-dual form for the potential of the
quintessence field,
\begin{equation}
V(\phi) = \Lambda \ {\rm sech}
(\kappa \, \phi/M_p) \,,
\label{Vsdual}
\end{equation}
where $\kappa$ is an order one parameter and $\Lambda \sim 10^{-120}
M_p^4$ is the cosmological constant. Actually, it is natural to take $\kappa =
\sqrt{2}$, and therefore it is not an extra parameter of the model as
it saturates the TCC bound.\footnote{We note in passing that a
  connection between the TCC and DESI DR2 results has been recently
  made evident in~\cite{Brandenberger:2025hof}.}

\begin{figure*}[htb!]
\begin{minipage}[t]{0.48\textwidth}
  \postscript{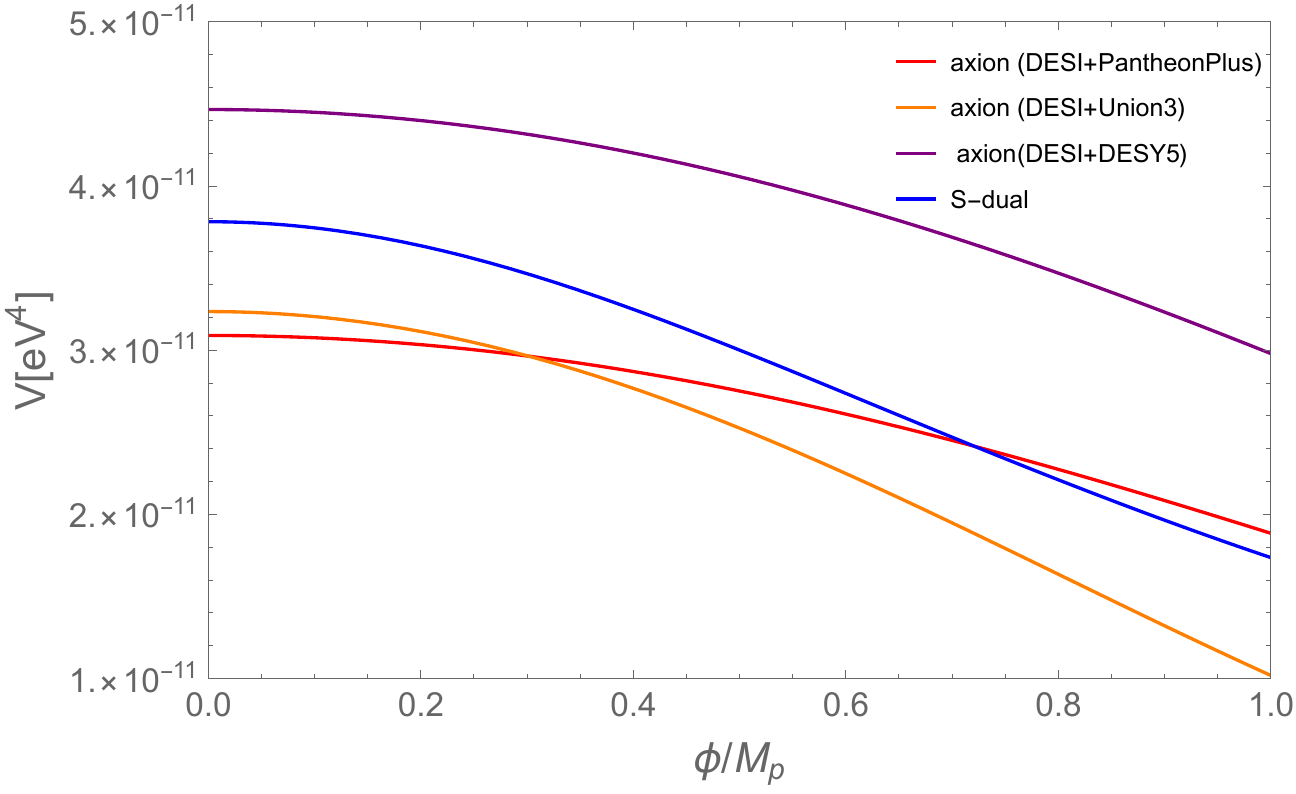}{0.96}
  \end{minipage}
\begin{minipage}[t]{0.49\textwidth}
    \postscript{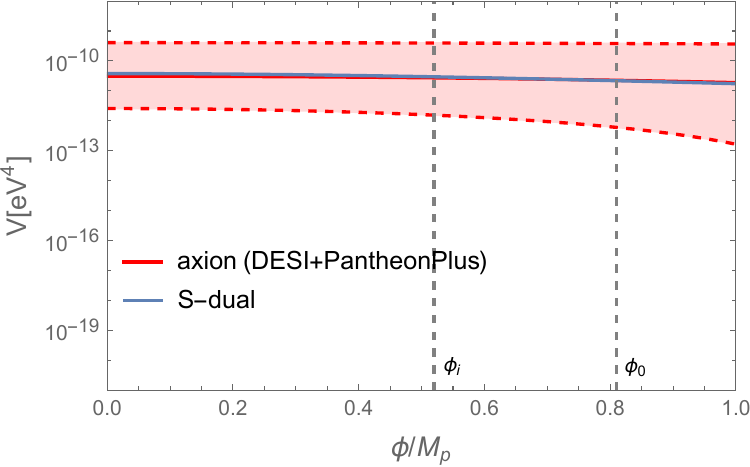}{0.9}
  \end{minipage}
\begin{minipage}[t]{0.50\textwidth}
    \postscript{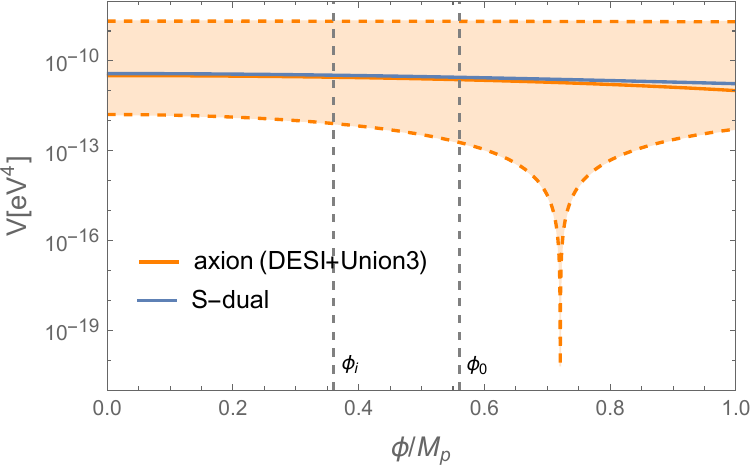}{0.9}
  \end{minipage}
  \begin{minipage}[t]{0.49\textwidth}
    \postscript{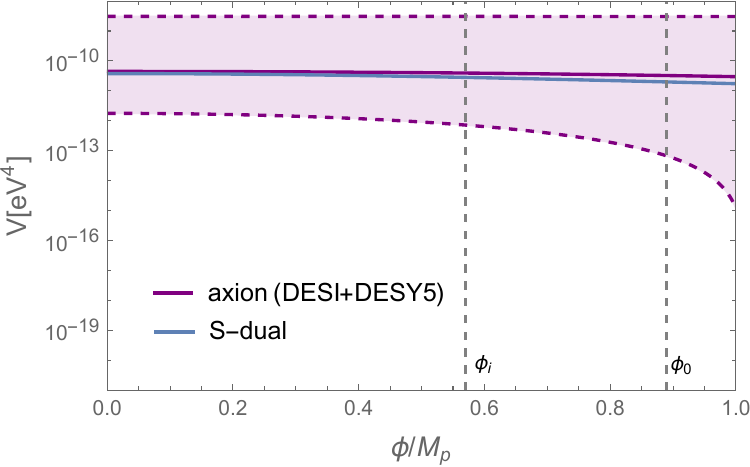}{0.91}
  \end{minipage}
  \caption{Comparison of the $S$-dual potential and the axion-like
    potential. The upper left panel shows a comparison between the $S$-dual
    potential and the axion-like potential with model parameters evaluated at the central
    values of the fit, for the different SN compilations. The
    other panels show a comparison between the $S$-dual potential and
    the $1\sigma$ region of the axion like potential for the different
    SN compilations: PantheonPlus (upper right), Union3 (lower left),
    and DESY5 (lower right). The shaded areas bounded by the dashed
    lines indicate the marginalized $1\sigma$ errors. The initial and
    present-day values of $\phi$ are indicated by the vertical dashed
    lines.
   \label{fig:1}}
\end{figure*}

Next, in line with our stated plan, we focus attention on the intriguing 
results recently reported by the DESI
Collaboration~\cite{DESI:2025zpo,DESI:2025zgx}. Measurements of baryon acoustic oscillations (BAO) from the second data
release (DR2) of the Dark Energy Spectroscopic Instrument
(DESI) have strengthened hints that dark energy may be weakening over time, casting some doubts on $\Lambda$ cold dark matter ($\Lambda$CDM)
cosmology~\cite{DESI:2025zpo,DESI:2025zgx}. A major challenge for $\Lambda$CDM is an epoch dependent
fit to the equation of state. More concretely, the DESI Collaboration
analyzed the $(w_0,w_a)$ plane, assuming that the equation of state of
dark energy satisfies
\begin{equation}
  w(z) = w_0 + w_a \ \frac{z}{(1 + z)} \,,
\end{equation}
as a function of the redshift $z$. The constraints on the parameters
using DESI DR2 BAO data alone are rather weak,
\begin{equation}
  \left. \begin{array}{ccl}
           w_0 & = & -0.48^{+0.35}_{-0.17}\\
           w_a & < & -1.34 \\
           \end{array} \right\} {\rm DESI \ BAO} \,,
\end{equation}
but they define a degeneracy direction in the
$(w_0,w_a)$ plane, though these constraints do not show a strong preference for
dark energy evolution~\cite{DESI:2025zgx}. Now, when DESI DR2 data are
combined with information from the cosmic microwave background (CMB)~\cite{Planck:2018vyg}
and supernova (SN) datasets (PantheonPlus~\cite{Brout:2022vxf}, Union3~\cite{Rubin:2023ovl}, and DESY5~\cite{DES:2024jxu}) the likelihood analysis shows evidence for
a time-evolving dark energy equation of state, yielding the following marginalized posterior results:
\begin{equation}
  \left. \begin{array}{ccl}
           w_0 & = & -0.838 \pm 0.055\\
           w_a & = & -0.62^{+0.22}_{-0.19} \\
           \end{array} \right\} \begin{array}{l}{\rm DESI+CMB}\\{\rm +
                                  PantheonPlus}\\
                                  \end{array}\,,
       \end{equation}

       \begin{equation}
  \left. \begin{array}{ccl}
           w_0 & = & -0.667 \pm 0.088\\
           w_a & = & -1.09^{+0.31}_{-0.27} \\
         \end{array} \right\}
\begin{array}{l}{\rm DESI+CMB}\\{\rm +
                                  Union3}\\
                                  \end{array} \,,
       \end{equation}
       
\begin{equation}
  \left. \begin{array}{ccl}
           w_0 & = & -0.752 \pm 0.057\\
           w_a & = & -0.89^{+0.23}_{-0.20} \\
         \end{array} \right\}
\begin{array}{l}{\rm DESI+CMB}\\{\rm +
                                  DESY5}\\
                                  \end{array}\,;
\end{equation}
respectively~\cite{DESI:2025zgx}. The preference for
  dynamical dark energy remains robust
when confronting DESI data with other well-known parameterizations of
$w(z)$~\cite{Giare:2024gpk}.

The DESI+CMB+SN constraints have an
unambiguous preference for a sector of the $(w_0,w_a)$ plane in which
$w_0 > -1$ and $w_0 + w_a < -1$, suggesting that $w(z)$ may have experienced a transition from a phase violating the null
energy condition at large $z$ to a phase obeying it at small
$z$. As shown in~\cite{Shlivko:2024llw},  this impression is misleading, because rather
simple quintessence models satisfying the null energy condition for all $z$, predict an
observational preference for the same sector; see
also~\cite{Abreu:2025zng,Shajib:2025tpd} for additional
examples.\footnote{Additionally, it is noteworthy that if a quintessence scalar field (with
positive kinetic energy) couples to dark matter, the effective
equation of state parameter $w_{\rm eff}$ could transition from
$w_{\rm eff} < -1$ in
the distant past to $w_{\rm eff} > - 1$ in the present epoch~\cite{Chakraborty:2025syu}. This
construct accommodates the region of the prameter space favored by the
DESI data, but the dark energy sector is devoid of any
pathologies with the equation of state parameter $w_\phi > -1,$ $\forall
\phi$. The preference of DESI data for an interacting dark sector was first noted in~\cite{Giare:2024smz}.} Of  
particular interest here, the hilltop quintessence model with
axion-like potential is given by, 
\begin{equation}
  V (\phi) = m_a^2 \ f_a^2 \ [ 1 + \cos(\phi/f_a)] \,,
\label{Vaxion}
\end{equation}
where $m_a$ denotes the mass of the boson particles related to the
scalar field and $f_a$ is regarded as the effective energy~\cite{Freese:1990rb,Frieman:1995pm}. Using DESI
DR2 results, CMB observations, and the three SN datasets the DESI
Collaboration reported the following constraints on the physical mass
and the effective energy scale: 
\begin{itemize}[noitemsep,topsep=0pt]
\item $\log(m_a/{\rm eV})
= -32.67^{+0.23}_{-0.25}$ for PantheonPlus, $-32.50^{+0.28}_{-0.30}$
for Union3, and $-32.63^{+0.26}_{-0.30}$ for DESY5;
\item  $\log(f_a/M_p) = -0.13^{+0.33}_{-0.29}$ for PantheonPlus, $-0.29^{+0.63}_{-0.35}$ for Union3, and
$-0.09^{+0.66}_{-0.40}$ for 
DESY5;
\end{itemize}
respectively~\cite{Lodha:2025qbg}. The constraints demand that the field starts in the
hilltop regime, with initial condition 
\begin{equation}
\phi_i/f_a \sim 0.7 \, .
\label{phii}
\end{equation}
Then, the field rolls down the potential, reaching the present value of
$\phi_0/f_a \sim 1.1$, traversing approximately $\Delta \phi
\sim 0.4 M_p$. A point worth noting at this juncture is that (\ref{phii}) is a result of a multi-parameter space likelihood analysis using the potential  (\ref{Vaxion}), with a prior on initial conditions of $\phi_i/f_a > 0.01$~\cite{DESI:2025hce}.

A genuine concern about the $S$-dual quintessence model would be that
the assumption that the field starts close to the top of the potential
may be an unnatural fine-tuning. However, it was recently pointed out
in~\cite{Chen:2025rkb} that if the top of the hill corresponds to an
enhanced symmetry point, then the requirement for the scalar field to
start rolling near a local maximum in the potential may not be
unnatural. This is because a top of the hill potential with an
enhanced gauge symmetry point (e.g., for the
$S$-dual potential, a $\mathbb{Z}_2$ symmetric point) corresponds to spontaneous breaking of the gauge
symmetry as we roll away. We can then envision that the universe could
have started at a higher temperature, where the generic expectation is
that the symmetries get restored, and the symmetric point would be
emerging as a preferred point. As the universe cools off the symmetry
gets broken leading to a natural rolling away from the symmetric point
and the start of late-time cosmology. Quantum fluctuations will move
the field away from the top. Such fluctuations are, however, very small
(of order $\phi_i \sim H \sim 10^{-60}$), and it takes $t \sim 1/H \
\ln (M_p/H)$ about 100 times more than the age of the universe for the
field $\phi$ to reach values of order
unity~\cite{Vafa:2025nst,Rudelius:2019cfh}. Thus, presumably some thermal
fluctuation could provide the required deformation to
accommodate the initial condition given in (\ref{phii}).\footnote{For a different viewpoint on initial
  conditions on hilltop potentials,
  see~\cite{Antoniadis:2020bwi,Cicoli:2021skd,Bhattacharya:2024hep,Cicoli:2024yqh,Bhattacharya:2024kxp,Borghetto:2025jrk}.}

In Fig.~\ref{fig:1} we compare the
    $S$-dual potential of (\ref{Vsdual})
    with $\kappa = \sqrt{2}$ and the axion-like potential of
(\ref{Vaxion}) for the different SN datasets. As one can check by inspection, the-$S$ dual potential
    with the choice of initial condition  and $\kappa = \sqrt{2}$ is almost indistinguishable from the axion-like potential,
    independently of the SN dataset.

\begin{figure}[htb!]
   \begin{minipage}[t]{0.49\textwidth}
    \postscript{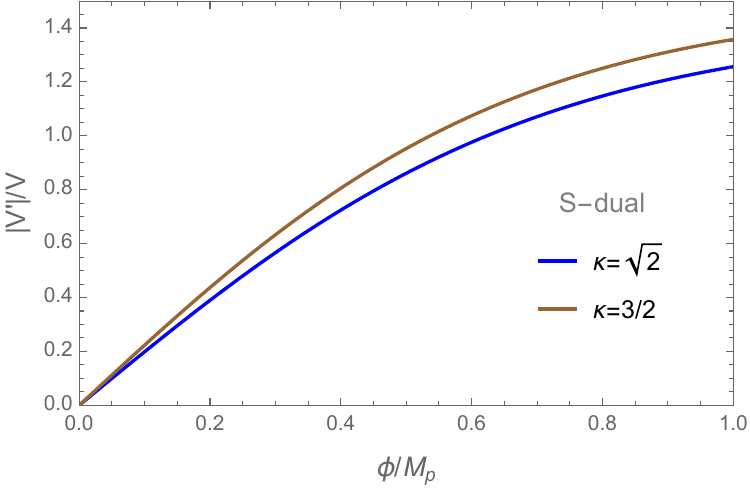}{0.9}
  \end{minipage} \\
\begin{minipage}[t]{0.49\textwidth}
    \postscript{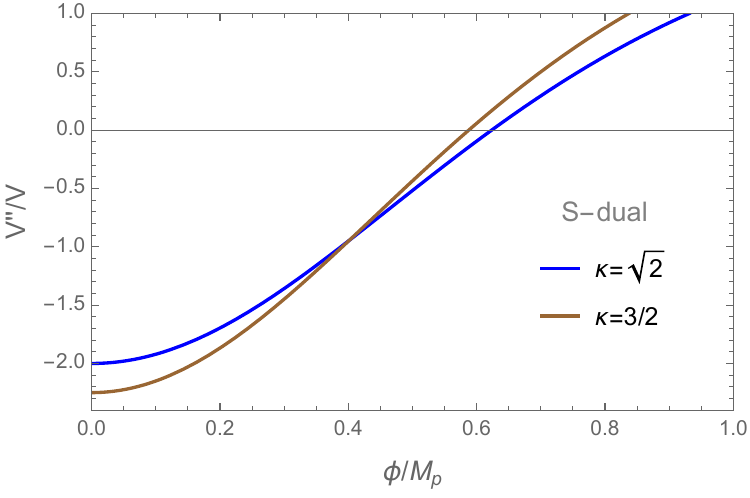}{0.9}
  \end{minipage}\\
  \begin{minipage}[t]{0.49\textwidth}
    \postscript{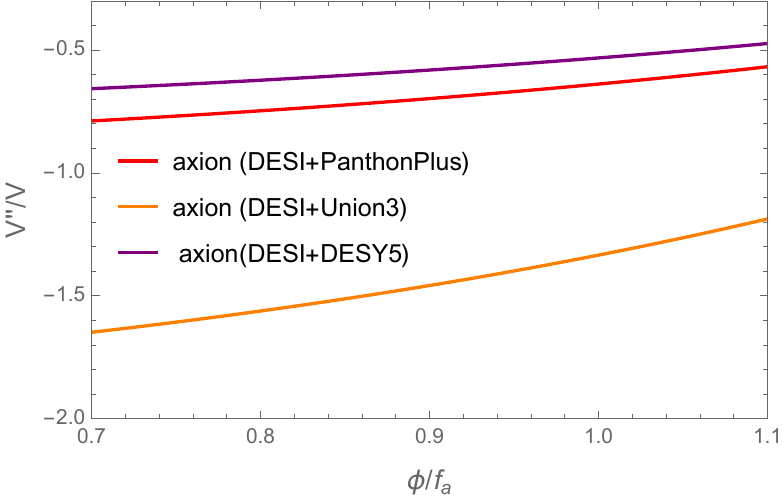}{0.9}
  \end{minipage}
  \caption{dS conjecture: $|V'|/V$ (upper) and $V''/V$ (middle) for
    the $S$-dual potential, and $V''/V$ (lower) for the axion-like 
    potential. \label{fig:2}}
\end{figure}

\begin{figure}[t]
  \vspace{0.5cm}
    \postscript{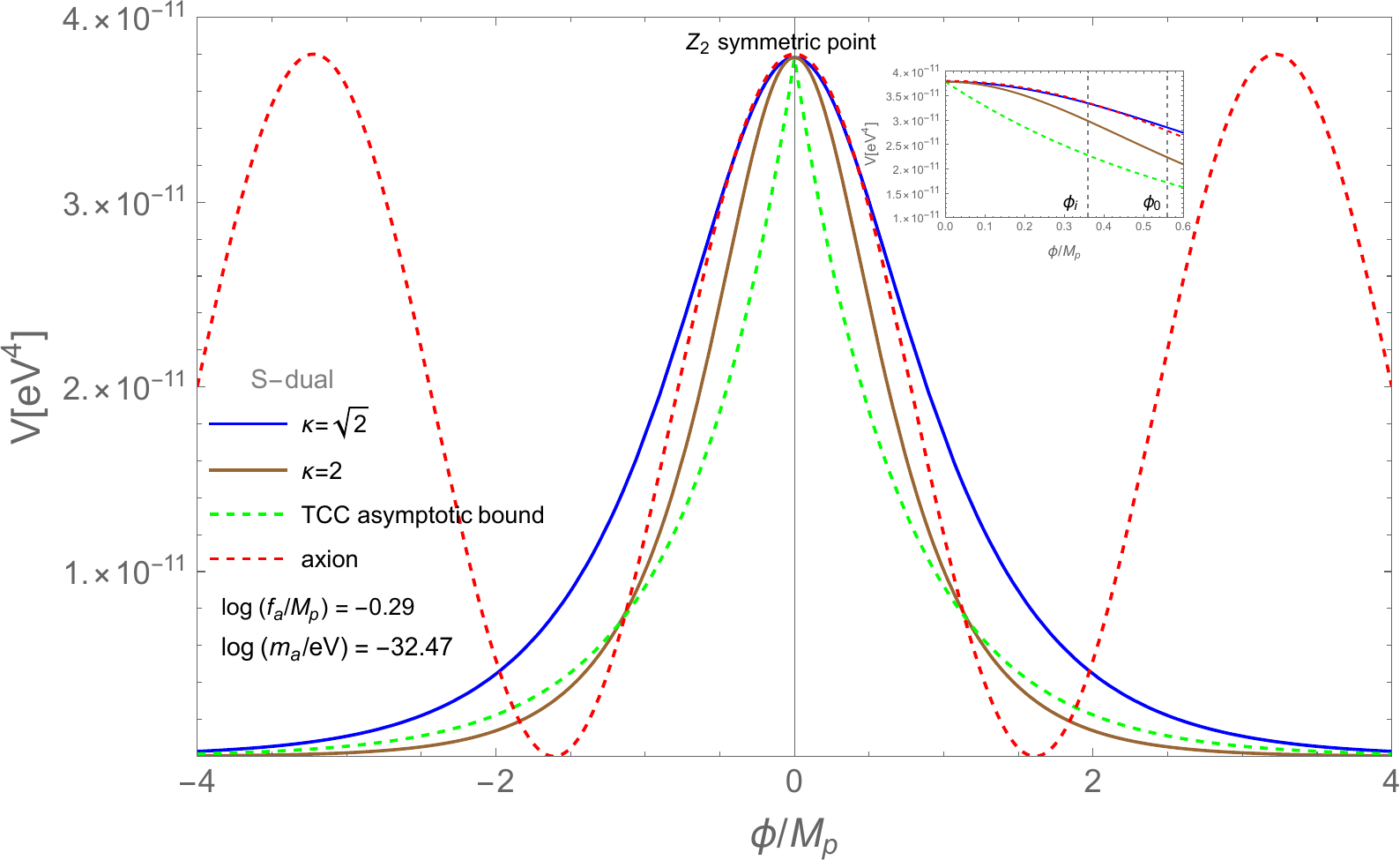}{0.98}
    \caption{Comparison of the $S$-dual potential and the axion-like
      potential for specific choice $m_a$ and $f_a$, which are
      consistent with DESI DR2 results for all SN datasets. The TCC
      asymptotic bound is indicated by the dashed green line. The
      $S$-dual potential with $\kappa = \sqrt{2}$ asymptotes to the
      TCC bound. The case
      with $\kappa = 2$ can be taken as an illustration of the
      uncertainty associated with $f_a$.
      \label{fig:3}}
    \end{figure}

The last item in the agenda is to verify whether the $S$-dual and
axion-like potentials satisfy the dS conjecture. In Fig.~\ref{fig:2}
we show $|V'|/V$ for the $S$-dual potential as well as $V''/V$ for the
axion and $S$-dual potentials, where $V' \equiv \partial V/\partial
\phi$. For the purposes of this investigation, we
can ignore the criterium (\ref{c2}), because we are considering the
specific application of hilltop quintessence scalar fields as models for dark
energy~\cite{Agrawal:2018rcg,Storm:2020gtv}. For the
    $S$-dual potential (\ref{Vsdual}), we explore the dS conjecture
    for both $\kappa =\sqrt{2}$ and
    $\kappa = 3/2$. The latter, which is also consistent with the TCC bound,
    can be taken as an illustration of the uncertainty associated with
    $f_a$. For a thorough discussion of the TCC bound on
      the axion-like potential, see~\cite{Shlivko:2023wkx}.  Note that both (\ref{Vsdual}) and (\ref{Vaxion}) satisfy
    the dS conjecture, independently of  the SN dataset selection.  

In summary, we have proposed a dark energy model in which a quintessence field rolls near the vicinity of a local 
maximum of its potential characterized by (\ref{Vsdual}). We
confronted the model with Swampland ideas and showed that the potential
is consistent with the distance conjecture, the de Sitter conjecture,
and the TCC. We also investigated the compatibility of (\ref{Vsdual})
with the intriguing DESI DR2 findings and demonstrated that the shape
of the S-dual potential is almost indistinguishable from the
axion-like potential (\ref{Vaxion}) analyzed by the DESI
Collaboration. Our results are encapsulated in Fig.~\ref{fig:3}.
 We have taken the natural choice of $\kappa = \sqrt{2}$, which makes
 $\kappa$ not an extra free parameter in the model, as it saturates
the TCC bound. This implies that $S$-dual quintessence cosmology
eliminates one free parameter compared to
the axion-like potential (\ref{Vaxion}) analyzed by the DESI
Collaboration. Furthermore, in our $S$-dual model $\phi= 0$ is an
enhanced gauge symmetry point, {\it viz.} a $\mathbb{Z}_2$ symmetric
point. This guarantees that $V (\phi)$ is critical at $\phi=0$, and so
in consonance with~\cite{Chen:2025rkb}  
the idea that the scalar field is rolling near a local maximum in the potential
 is natural by $\mathbb{Z}_2$ gauge symmetry, which gets broken
 at low temperature. All in all, our $S$-dual quintessence model has
 only one free parameter, 
and can be confronted with future cosmological observations.

\section*{Acknowledgements}

We are thankful to Cumrun Vafa for valuable discussions and insightful remarks on
the first draft of this letter. We thank Edward Witten for an
interesting comment on the manuscript. The work of L.A.A. is supported by the U.S. National Science
Foundation (NSF Grant PHY-2412679). I.A. is supported by the Second
Century Fund (C2F), Chulalongkorn University.  The work of D.L. is supported by the Origins
Excellence Cluster and by the German-Israel-Project (DIP) on
Holography and the Swampland.

\end{document}